\documentclass[aps,pra,10pt,showpacs,twocolumn,superscriptaddress,longbibliography]{revtex4-1}
\usepackage{physics}
\usepackage{graphicx}
\usepackage[usenames]{color}
\usepackage{amssymb,amsmath}
\usepackage{siunitx}%To use degree symbol
\usepackage{xcolor}
\usepackage{setspace} 
\usepackage{float} 
\usepackage{tabularx}
\usepackage[linkcolor=blue,citecolor=blue,colorlinks=true,urlcolor=blue]{hyperref}
\newcommand{\eq}[1]{Eq.~(\ref{#1})}
\newcommand{\fig}[1]{Fig.~\ref{#1}}
\newcommand{\be}[1]{\begin{equation}\label{#1}}
\newcommand{\ee}{\end{equation}}
\usepackage{amsmath,esint}
\usepackage{comment}
\usepackage{lipsum}
\usepackage[toc]{appendix}
\usepackage{hyperref}
\usepackage{multirow}
\hypersetup{
    colorlinks=false, %set true if you want colored links
    linktoc=all,     %set to all if you want both sections and subsections linked
    linkcolor=blue,  %choose some color if you want links to stand out
}

\begin{document}

\title{Formation of singly ionized oxygen atoms from O$_2$ driven by XUV pulses: a toolkit for the break-up of FEL-driven diatomics}

%\title{Formation of singly-ionized oxygen atoms during the breakup of O$_2$ driven by an XUV pulse}

\author{M. Mountney}
\affiliation{Department of Physics and Astronomy, University College London, Gower Street, London WC1E 6BT, United Kingdom}
\author{Z. Wang}
\affiliation{Department of Physics and Astronomy, University College London, Gower Street, London WC1E 6BT, United Kingdom}
\author{F. Trost}
\affiliation{Max-Planck-Institut für Kernphysik, Heidelberg, Germany}
\author{H. Lindenblatt}
\affiliation{Max-Planck-Institut für Kernphysik, Heidelberg, Germany}
\author{A. Magunia}
\affiliation{Max-Planck-Institut für Kernphysik, Heidelberg, Germany}
\author{R. Moshammer}
\affiliation{Max-Planck-Institut für Kernphysik, Heidelberg, Germany}
\author{T. Pfeifer}
\affiliation{Max-Planck-Institut für Kernphysik, Heidelberg, Germany}
\author{A. Emmanouilidou}
\affiliation{Department of Physics and Astronomy, University College London, Gower Street, London WC1E 6BT, United Kingdom}

\begin{abstract}
We formulate a general hybrid quantum-classical technique to describe the interaction of diatomic molecules with XUV pulses. We demonstrate the accuracy of our model in the context of the interaction of the O$_2$ molecule with an XUV pulse with photon energy ranging from 20 eV to 42 eV. We account for the electronic structure and electron ionization quantum mechanically employing accurate molecular continuum wavefunctions. We account for the motion of the nuclei using classical equations of motion. However, the force of the nuclei is computed by obtaining accurate potential-energy curves 
 of O$_2$ up to O$_2^{2+}$, relevant to the 20 eV-42 eV photon-energy range,  using advanced quantum-chemistry techniques. We find the dissociation limits of these states and the resulting atomic fragments and employ   the Velocity Verlet algorithm to compute the velocities of these fragments. We incorporate both electron ionization and nuclear motion in a stochastic Monte-Carlo simulation and identify the ionization and dissociation pathways when O$_2$ interacts with an XUV pulse. Focusing on the O$^+$ + O$^+$ dissociation pathway, we obtain the kinetic-energy release distributions of the atomic fragments and find very good agreement with experimental results. Also, we explain the main features of the KER in terms of ionization sequences consisting of two sequential single-photon absorptions resulting in different  O$^+$ and O$^{2+}$ electronic state configurations involved in the two transitions.\end{abstract}

\maketitle

\section{Introduction}
In recent years, the advent of extreme-ultraviolet (XUV) laser sources has opened up new avenues for probing ultrafast complex processes in molecular systems with unprecedented temporal and spatial resolution \cite{ref:Zhao,ref:Dudovich}. In particular, experiments utilise free-electron lasers (FELs), such as FLASH in Hamburg \cite{ref:Ackermann,ref:Faatz}, to produce intense XUV pulses. The high photon energy of XUV radiation results in phenomena such as multi-electron ionization and dissociation in molecules. A large number of ionization sequences can contribute to a certain dissociation pathway in strongly driven molecules. This results in a plethora of features in the sum of the kinetic energies of the atomic fragments, i.e. the kinetic energy release (KER), as a function of photon energy of the atomic fragments resulting from molecular dissociation \cite{ref:Magrakvelidze,ref:Rudenko}. Hence, using advanced theoretical models to identify the ionization sequences leading to molecular dissociation is crucial for understanding experimental results in physical, chemical and biological processes taking place when molecules interact with XUV pulses \cite{ref:Lin,ref:Nisoli,ref:Miller,ref:Cattaneo,ref:Borrego}.

Molecular oxygen, O$_2$, is of great interest due to its significance in biology and atmospheric chemistry. Specifically, oxygen  is the key component in both the metabolic processes of all living organisms and the ozone layer which shields us from the Sun's ultraviolet radiation \cite{ref:Schmidt-Rohr,ref:Jensen,ref:Paterson,ref:Parker}. There has already been a number of studies on singly and doubly ionized O$_2$, for instance see Refs \cite{ref:Corlin,ref:Schmid,ref:Magunia}. Here, we formulate a hybrid quantum-classical theoretical model to address the interaction of molecules with intense pulses of high photon energy. 
 We focus on  the formation of two O$^+$ atomic fragments in coincidence resulting from the interaction of O$_2$ with an XUV pulse with photon energy ranging from 20 eV to 42 eV. 
  We demonstrate that the theoretically obtained kinetic-energy release spectra as a function of the photon energy for the  O$^+$ + O$^+$ dissociation pathway are in very good agreement with experimental results.

Our hybrid quantum-classical model allows us to identify the ionization sequences that lead to the formation of the two O$^+$ fragments. Namely, for photon energies from 20 eV to 42 eV, 
up to two sequential single-photon ionization processes can take place starting from the ground state of O$_2$. The electronic configuration of the ground state of O$_2$ ($X^3\Sigma_g^-$) is given by ($1\sigma_g^2$, $1\sigma_u^2$, $2\sigma_g^2$, $2\sigma_u^2$, $3\sigma_g^2$, $1\pi_{ux}^2$, $1\pi_{uy}^2$, $1\pi_{gx}^1$, $1\pi_{gy}^1$), with the two open $\pi_g$ orbitals resulting in a triplet spin state.
 These single-photon ionization processes include the removal of valence or inner-valence electrons from the ground state and singly excited states of O$_2$.  The different ionization sequences result from  the different O$_2$$^+$ states the molecule  transitions to after the removal of the first electron as well as the different 
 O$_2$$^{2+}$ states the molecule  transitions to after the escape of the second electron. For each sequence, we also identify the times and the inter-nuclear distance when ionization takes place. This detailed analysis allows us  to relate specific features of the KER spectra for the O$^+$ + O$^+$ dissociation pathway with certain ionization sequences. We also investigate the dependence of the KER features and their connection to ionization sequences on pulse duration and intensity.
 
 %CITATIONS for Prof. Moshammer and Prof. Pfeifer:

%XUV-IR pump-probe setup used to obtain KER of O$_2^+$ \cite{ref:Corlin}

%XUV pump XUV probe setup used to obtain results \cite{ref:Schmid}

%Temporally resolve the dissociation of a specific O$_2^+$ state interacting with XUV pulse \cite{ref:Magunia}

The hybrid model we develop in this work consists of employing state-of-the-art quantum-mechanical computations for describing electron escape to the continuum. At the same time, we allow the nuclei to move using classical equations of motion. 
 However, in these classical equations of motion the force between the nuclei   is provided by  employing accurate quantum-mechanically obtained potential-energy curves (PECs). Specifically, in order to calculate the necessary ionization cross sections for our theoretical calculations, we utilise state-of-the-art ab-initio quantum-mechanical techniques \cite{ref:Mountney_NO}. Namely, we compute the continuum wavefunction of the escaping electron using molecular orbital wavefunctions obtained in the Hartree-Fock framework \cite{ref:Agapi}. Most other studies do not provide such accurate molecular wavefunctions and instead  use approximations that employ  linear combination of atomic orbitals. Also, to provide the force in the classical equations that account for the motion of the nuclei, we obtain accurate PECs of the various states of O$_2$ and singly and doubly ionized O$_2$ by using advanced quantum-mechanical techniques in the framework of the quantum-chemistry package MOLPRO \cite{ref:Molpro,ref:Bhattacharya,ref:Hadjipittas}. Computing PECs of these various states is not an easy task due to the open-shell configuration of O$_2$, especially in the cases where an electron is removed from an inner-valence orbital.

In Section \ref{sec:Method}, we discuss our theoretical methods for describing the interaction of a diatomic molecule, in this work O$_2$, with an XUV pulse. In Section \ref{sec:Experiment}, we outline the experimental set up for the same interaction. Then, in Section \ref{sec:Results}, we discuss our results for the KER of the two O$^+$ fragments. In particular, we consider both a low- and high-intensity XUV pulse in the photon-energy range from 20 eV to 42 eV. We identify the key features of the KER distributions, as well as the ionization sequences that lead to the obtained spectra.

\section{Theoretical method}\label{sec:Method}
In what follows, we formulate a hybrid quantum-classical model to describe the interaction of  a diatomic molecule, in this work O$_2$,  with an XUV pulse. This model adapts the Born-Oppenheimer approximation \cite{ref:Born} in order to separate the nuclear from the electronic motion in the driven molecule. We model the ionization of electrons quantum mechanically, as described in Section \ref{ssec:Electronic}. We account for the motion of the nuclei and compute the velocities of the atomic fragments using the classical equations of motion of the two-body system, see Section \ref{ssec:Nuclear}.  In   Section \ref{ssec:VV}, we describe how we use the Velocity-Verlet algorithm to calculate the internuclear distance and momentum at each time step. The force in these classical equations is computed via the PECs of up to doubly ionized Oxygen, which are obtained using advanced quantum-chemistry techniques, see Section \ref{ssec:PECs}.  In Section \ref{ssec:Wigner}, we describe how we sample the initial conditions for the nuclei using the Wigner distribution for the Morse oscillator. 
 Both electron escape and nuclear motion are incorporated in a Monte-Carlo simulation described in Section \ref{ssec:MC}. In 
 Section \ref{ssec:Diss}, we outline the criteria we use to identify the time in the Monte-Carlo simulation when  the molecule  dissociates transitioning to two atomic fragments interacting with an XUV pulse.
\subsection{Single-photon ionization cross sections}\label{ssec:Electronic}
To model ionization, we obtain the quantum wavefunctions for the bound and continuum electrons. We calculate wavefunctions of the bound electrons with the Hartree-Fock (HF) method employing the quantum-chemistry package MOLPRO \cite{ref:Molpro}. When using MOLPRO, we employ the augmented Dunning correlation consistent quadruple valence basis set (aug-cc-pVQZ) \cite{ref:Dunning}. Using these bound electron wavefunctions, we then solve a system of HF equations \cite{ref:Mountney,ref:Agapi} to obtain the wavefunction of an electron that escapes to the continuum after absorbing a single photon. We express the continuum and bound state wavefunctions using the single center expansion (SCE) \cite{ref:Demekhin,ref:Agapi} 
\begin{equation}
\psi(\boldsymbol{r}) = \sum_{lm}\frac{P_{lm}(r)Y_{lm}(\theta, \phi)}{r},
\label{eqn:SCE}
\end{equation}
where $l,m$ are the angular momentum and magnetic quantum numbers respectively, $Y_{lm}$ is a spherical harmonic and $P_{lm}$ is the radial part of the wavefunction. Note that we fully account for the Coulomb potential when solving to obtain the continuum wavefunctions.

We use the continuum and bound wavefunctions to calculate the photoionization cross sections for an electron transitioning from an initial orbital $\psi_i$ to a final continuum orbital $\psi_{\epsilon}$. These cross sections are given by \cite{ref:Banks,ref:Sakurai}
\begin{equation}
\label{eqn:Xsec}
\sigma_{i\rightarrow\epsilon} = \frac{4}{3}\alpha\pi^2\omega N_i \sum_{M=-1,0,1}\abs{D^M_{i\epsilon}}^2,
\end{equation}
where $\alpha$ is the fine-structure constant, $\omega$ the photon energy, $N_i$ the occupation number of orbital $i$ and $M$ the polarization of the photon. In the length gauge, the dipole matrix element, $D^M_{i\epsilon}$, is given by the following
\begin{equation}
\label{eqn:DM}
D^M_{i\epsilon} = \int \psi^*_{\epsilon}(\boldsymbol{r})\psi_{i}(\boldsymbol{r})\sqrt{\frac{4\pi}{3}}Y_{1M}(\theta,\phi)d\boldsymbol{r}.
\end{equation}
Substituting \eq{eqn:SCE} in \eq{eqn:DM} and integrating over the angular components, we obtain the following expression in terms of Wigner-3$j$ symbols and in terms of  the radial bound and continuum wavefunctions, $P_{l_im_i}$ and $\mathcal{P}_{l'm'}$, respectively
\begin{equation}
\begin{split}
\label{eqn:Dl1m1miWigner}
    D^M_{i\epsilon} &= \sqrt{\frac{4\pi}{3}}\sum_{l',m',l_i,m_i}(-1)^{m'}\sqrt{(2l_i+1)(2l'+1)}
    \\
    &\times\begin{pmatrix}
    l' & l_i & 1\\
    0 & 0 & 0 \\
    \end{pmatrix}
    \begin{pmatrix}
    l' & l_i & 1\\
    -m' & m_i & M \\
    \end{pmatrix}
    \\
    &\times
    \int^{\infty}_0 dr \mathcal{P}_{l'm'}(r)rP_{l_im_i}(r).
\end{split}
\end{equation}
Note that we also include Auger-Meitner decay in our calculations. However, for the photon energy range of 20 eV to 42 eV only a very small number of Auger-Meitner transitions are energetically allowed. Hence, in this work, we do not discuss the computation of Auger-Meitner rates which for diatomic  molecules are given in Ref. \citep{ref:Agapi}.

\subsection{Two-body equations for nuclear motion}\label{ssec:Nuclear}
In classical mechanics the Lagrangian of a two-body system with potential energy $U(\boldsymbol{r})$ is given by
\begin{equation}
L = \frac{1}{2}m_1\dot{\boldsymbol{x}}_1^2 + \frac{1}{2}m_2\dot{\boldsymbol{x}}_2^2-U(\boldsymbol{r}),
\label{eqn:TwoBody}
\end{equation}
where $m_1, m_2$ and $\dot{\boldsymbol{x}}_1,\dot{\boldsymbol{x}}_2$ are the masses and velocities of the two bodies, respectively. In our case, the two bodies are the nuclei of the diatomic molecule. The potential is a function of the difference between the position vectors of the nuclei, namely the difference vector 
$\boldsymbol{r} = \boldsymbol{x}_1 - \textbf{x}_2$. The internuclear distance of the molecule is $r = \abs{\boldsymbol{r}}$. The center of mass, $\boldsymbol{R}$, of the molecule is defined as
\begin{equation}
\begin{split}
\boldsymbol{R} &= \frac{m_1}{m_1+m_2}\boldsymbol{x}_1 + \frac{m_2}{m_1+m_2}\boldsymbol{x}_2
\\
&= \frac{\mu}{m_2}\boldsymbol{x}_1 + \frac{\mu}{m_1}\boldsymbol{x}_2,
\end{split}
\end{equation}
where $\mu = \frac{m_1m_2}{m_1+m_2}$ is the reduced mass of the molecule. The Lagrangian in \eq{eqn:TwoBody} can be rewritten in terms of the center of mass velocity, $\dot{\boldsymbol{R}}$, and the relative velocity between the nuclei, $\dot{\textbf{r}}$ as follows \cite{ref:Goldstein}
\begin{equation}
L = \frac{1}{2}(m_1+m_2)\dot{\boldsymbol{R}}^2 + \frac{1}{2}\mu\dot{\boldsymbol{r}}^2-U(\textbf{r}).
\label{eqn:TwoBody2}
\end{equation}
By  Newton's second and third law, the force between the nuclei is given by
\begin{equation}
\label{eqn:Force}
F = -\frac{\partial H}{\partial r} = -\frac{dU}{dr}.
\end{equation}
The position and velocities of the nuclei can be expressed in terms of $\boldsymbol{r}$ and the reduced mass as follows
\begin{equation}
\label{eqn:TwoBody3}
\begin{split}
&\boldsymbol{x}_1 = \frac{\mu}{m_1}\boldsymbol{r} \implies \dot{\boldsymbol{x}}_1 = \frac{\mu}{m_1}\dot{\boldsymbol{r}}
\\
&\boldsymbol{x}_2 = -\frac{\mu}{m_2}\boldsymbol{r} \implies \dot{\boldsymbol{x}}_2 = -\frac{\mu}{m_2}\dot{\boldsymbol{r}}.
\end{split}
\end{equation}

\subsection{Algorithm for propagating the nuclei in time} \label{ssec:VV}
Next, we explain how we obtain the velocities of the nuclei $\dot{\boldsymbol{x}}_1$ and $\dot{\boldsymbol{x}}_2$. \eq{eqn:TwoBody3} shows that to obtain the final velocities, we need to track the internuclear distance, $r$, and the relative velocity, $\boldsymbol{v}=\dot{\boldsymbol{r}}$, as a function of time. To do so, we employ the Velocity-Verlet algorithm \cite{ref:Verlet}. This algorithm calculates recursively the internuclear distance and the magnitude of the relative velocity at each time step as follows
\begin{equation}
\begin{split}
&r_{n+1} = r_n + v_n\Delta t + \frac{1}{2\mu}F_n(\Delta t)^2
\\
&v_{n+1} = v_n + \frac{1}{2\mu}(F_{n+1}+F_n)(\Delta t),
\end{split}
\label{eqn:Verlet}
\end{equation}
where $\Delta t$ is the time step of the propagation and $F_n$ is the force at each time step given by
\begin{equation}
\begin{split}
F_n &= -\frac{dU(r_n)}{dr}
\\
F_{n+1} &= -\frac{dU(r_{n+1})}{dr},
\end{split}
\end{equation}
where $U(r_n)$ is the potential energy at the end of the n$^{\text{th}}$ time step, while $U(r_{n+1})$ is the potential energy at the end of the (n+1)$^{\text{th}}$ time step. To obtain the potential $U(r)$, we compute the PECs of the singly and doubly ionized states of O$_2$. We describe how to do so in the next section. Hence, obtaining $v$ at each time step allows us, through \eq{eqn:TwoBody3}, to compute the magnitudes of the velocities $\dot{x}_1$ and $\dot{x}_2$ of the two atomic fragments.

\subsection{Computation of potential-energy curves} \label{ssec:PECs}
 Next, we describe the computation of the PECs of O$_2$ up to O$_2^{2+}$ ionic states. We perform these calculations for all singly and doubly ionized states of O$_2$ that are energetically accessible by absorption of a single photon with energy varying from 20 eV to 42 eV. This range of photon energies suffices to ionize electrons from all outer-valence orbitals and the inner-valence orbital $2\sigma_g$. As a result, we can access and thus compute  the PECs of eighteen states which are comprised of the ground state of O$_2$, five O$_2^+$ and twelve O$_2^{2+}$ states.

First, we employ the HF method in MOLPRO to obtain, as a function of distance, the bound orbitals of O$_2$ in its ground state. We use these orbitals as input in the subsequent computation of the PECs employing the complete active space self-consistent field (CASSCF) method \cite{ref:Werner_CASSCF, ref:Knowles_CASSCF}. In the current work, in a similar fashion to the one followed in Ref. \cite{ref:Bhattacharya} for the N$_2$ molecule, we consider 12 active orbitals. These orbitals include the nine occupied ones in the ground state of O$_2$, which are given by $1\sigma_g$, $1\sigma_u$, $2\sigma_g$, $2\sigma_u$, $3\sigma_g$, $1\pi_{ux}$, $1\pi_{uy}$, $1\pi_{gx}$, $1\pi_{gy}$ and the three virtual orbitals $3\sigma_u$, $4\sigma_g$ and $4\sigma_u$. Lastly, to obtain even more accurate PECs, using as input the bound orbitals obtained with the CASSCF method, we next employ the multireference configuration interaction (MRCI) method \cite{ref:Werner_MRCI, ref:Knowles_MRCI, ref:Knowles_MRCI2}. In our current computations, while the CASSCF method includes all possible excitations among only the 12 active orbitals, the MRCI method allows single and double excitations from all active orbitals to all orbitals. The MRCI method produces more accurate PECs compared to CASSCF by improving the description of the electron-electron repulsion \cite{ref:Werner_MRCI2,ref:Bhattacharya}. 

To compute the PECs, we utilise the same basis set as in Section \ref{ssec:Electronic}. For the states of singly and doubly ionized O$_2$ where at least one electron is missing from the inner valence orbitals $2\sigma_g$, there are several other states with the same symmetry and lower energy besides the desired state. This leads to variational collapse \cite{ref:Besley}, where MOLPRO computes the lowest-energy state with the same symmetry as the desired state. We address this issue during the calculations of the PECs with the CASSCF and MRCI methods by employing the state-averaging technique in MOLPRO. This technique ensures that MOLPRO computes a sufficient number of states with the same symmetry as a function of the internuclear distance. We identify the desired state by selecting the state that has the desired electronic configuration at the equilibrium distance of the ground state of the O$_2$ molecule. We find the equilibrium distance to be equal to $r_e = 1.205$ \AA \ = 2.28 a.u., in agreement with Refs. \cite{ref:Bytautas,ref:Weast}. As in Ref. \cite{ref:Bhattacharya}, to obtain the PECs of all states with electrons missing from outer-valence orbitals, we optimize all orbitals at the same time. We obtain the PECs for sixteen states using the MRCI states, while for the states O$_2^+(^4\Sigma_g^-)$ and O$_2^{2+}(^3\Sigma_g^-)$, we could not achieve convergence using the MRCI method and hence obtained the PECs using  the CASSCF method.

\begin{figure*}
	\centering
    \includegraphics[width=1\textwidth]{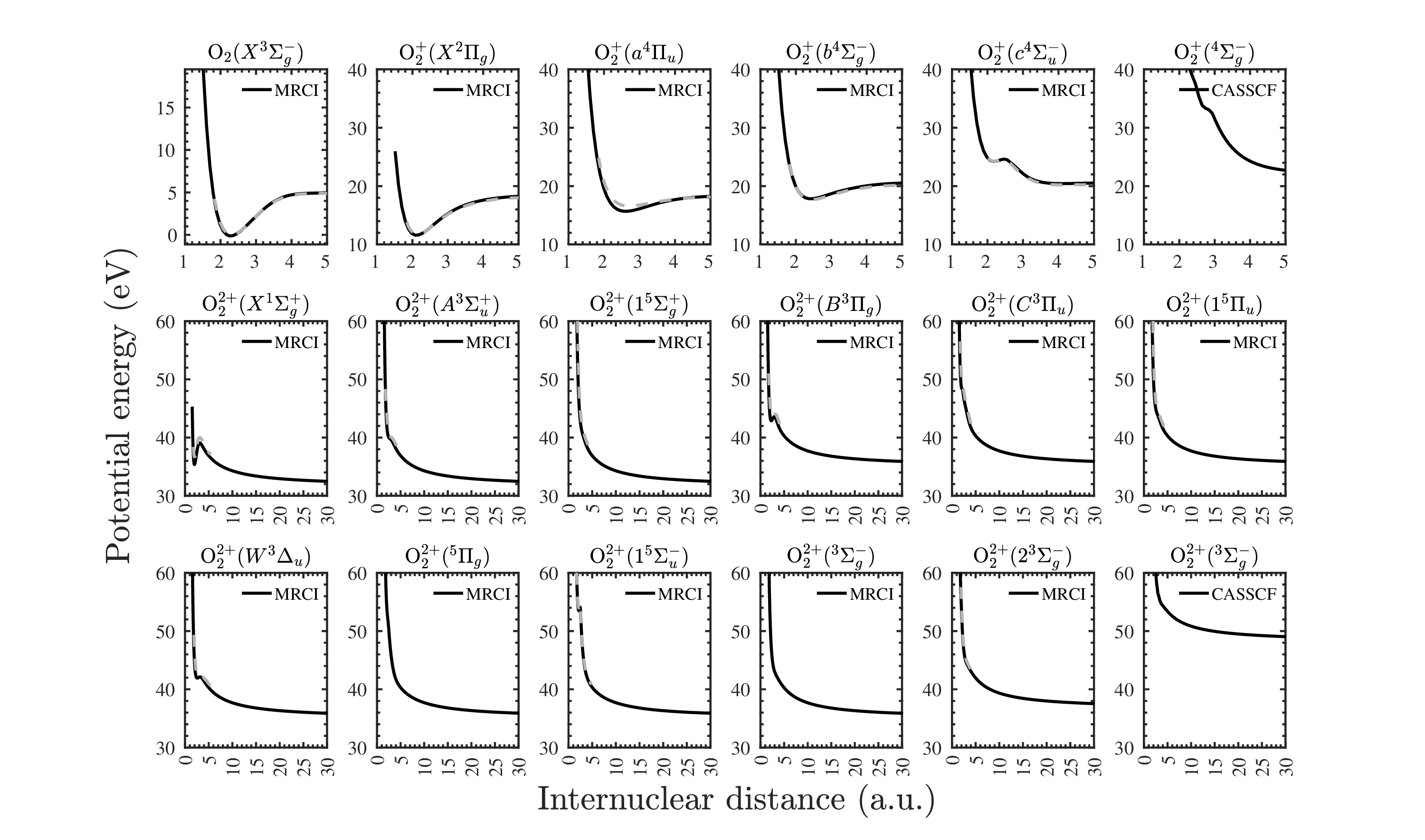}
\caption{Potential-energy curves of O$_2$, O$_2^+$ and O$_2^{2+}$ obtained using the techniques outlined in Section \ref{ssec:PECs} (black-solid line), compared with the PECs obtained in Refs. \cite{ref:Magrakvelidze,ref:Lundqvist,ref:Larsson} (gray-dashed lines).}
\label{fig:PECs}
\end{figure*}

In \fig{fig:PECs}, we compare the PECs we obtained using the method outlined above (black solid lines) with the theoretical PECs obtained in Refs.  \cite{ref:Magrakvelidze,ref:Lundqvist,ref:Larsson} (gray-dashed lines). The PECs obtained from the literature also utilise the MRCI method. In Refs.  \cite{ref:Magrakvelidze,ref:Lundqvist,ref:Larsson}, the PECs for O$_2^+(^4\Sigma_g^-)$, O$_2^{2+}(^5\Pi_g)$ and both of the O$_2^{2+}(^3\Sigma_g^-)$ states are not obtained, hence the lack of comparison in \fig{fig:PECs}, with our PECs for these states. To produce smooth PECs, we compute the potential energy of the states at internuclear distances in steps of $0.05$ \AA \ = 0.094 a.u. \fig{fig:PECs} shows that we find an excellent agreement for our computations of the PECs with the ones obtained in Refs.  \cite{ref:Magrakvelidze,ref:Lundqvist,ref:Larsson}. 

In the first row of \fig{fig:PECs}, we plot the PECs of the neutral ground state and the singly ionized states of O$_2$. These states possess a potential well and do not exhibit any repulsive behaviour, with the exception of the O$_2^+(^4\Sigma_g^-)$ state. In the second and third rows of \fig{fig:PECs}, we plot the PECs of the doubly ionized states of O$_2$. Due to the increased electrostatic repulsion between the ions, all of the O$_2^{2+}$ states exhibit repulsive behaviour. Note that only the O$_2^{2+}(X^1\Sigma_g^+)$, O$_2^{2+}(B^3\Pi_g)$ and O$_2^{2+}(W^3\Delta_u)$ states possess a potential well. However, the potential well of the O$_2^{2+}(W^3\Delta_u)$ state in particular is very shallow. To further verify the accuracy of the PECs we obtain in this work, we compare the atomic fragments resulting from the dissociation of each molecular state with the atomic fragments obtained in the literature \cite{ref:Magrakvelidze,ref:Hikosaka}. Next, we describe how to determine these atomic fragments.

\subsection{Dissociation of molecular states} \label{ssec:Diss}
To describe the interaction of O$_2$  with an XUV pulse, we use the Monte-Carlo technique described in Section \ref{ssec:MC}. If dissociation does take place, we need to specify the time during  propagation when the diatomic molecule dissociates  to two atomic fragments. We take this time to be when the energy of each molecular state, i.e. the PEC, converges to 99\% of each asymptotic value, known as  dissociation energy. From this time onwards, we perform Monte-Carlo simulations for the interaction of each of the two individual atomic fragments with the pulse. We calculate the dissociation energy for each molecular state by computing the potential energy at an internuclear distance equal to $10^4$ \AA.

Once we compute the dissociation energy of each molecular state of O$_2$ up to O$_2^{2+}$, we identify the atomic fragments resulting from the dissociation of each of these states. To do so, for each molecular state, we take all possible combinations of atomic fragments that sum to the same net charge as the molecular state under consideration. For all these combinations, we compute the sum of the energies of the atomic fragments and identify the one that matches the dissociation energy of the molecule. We compute the atomic energies with the MRCI method. The atomic fragments of each energetically accessible state of O$_2$ in the current work are given in Table \ref{table:DissFrags}. Moreover, in Table \ref{table:DissFrags} we provide the dissociation energies of the molecular states, which we compare with dissociation energies from the literature \cite{ref:Magrakvelidze,ref:Hikosaka}, and find them to be in very good agreement. For the sixteen states for which we were able to calculate their PECs using the MRCI method, we find a 1-2\% difference in the dissociation energies with respect to the literature. For the PECs where we were able to utilize only the CASSCF method, we find larger deviations of 7\% and 19\%. This is due to the difference in the techniques used. We also find agreement  with Refs. \cite{ref:Magrakvelidze,ref:Hikosaka} concerning the  atomic fragments resulting from the dissociation of the molecular states.

\begin{table*}[t]
\begin{tabular}{c|c|cc}
\multicolumn{1}{c|}{\multirow{2}{*}{ Molecular state }} & \multicolumn{1}{c|}{\multirow{2}{*}{ Atomic fragments }} & \multicolumn{2}{c}{Sum of energies of atomic fragments}                       \\ \cline{3-4} 
\multicolumn{1}{c|}{}  & \multicolumn{1}{c|}{} & \multicolumn{1}{c|}{ Our work (eV) } & \multicolumn{1}{c}{ Other work (eV) } \\ \hline
O$_2(X^3\Sigma_g^-)$ & O($^3P$)+O($^3P$) & \multicolumn{1}{c|}{5.0} & 5.0 \cite{ref:Magrakvelidze}  \\ 
O$_2^+(X^2\Pi_g)$ & O$^+$($^4$S)+O($^3$P) & \multicolumn{1}{c|}{18.5} & 18.8 \cite{ref:Magrakvelidze}\\   
O$_2^+(a^4\Pi_u)$ & O$^+$($^4$S)+O($^3$P) & \multicolumn{1}{c|}{18.5} & 18.8 \cite{ref:Magrakvelidze}\\   
O$_2^+(b^4\Sigma_g^-)$ & O$^+$($^4$S)+O($^1$D) & \multicolumn{1}{c|}{20.5} & 20.7 \cite{ref:Magrakvelidze}\\ 
O$_2^+(c^4\Sigma_u^-)$ & O$^+$($^4$S)+O($^1$D) & \multicolumn{1}{c|}{20.5} & 20.7 \cite{ref:Magrakvelidze}\\ 
O$_2^+(^4\Sigma_g^-)$ & O$^+$($^2$P)+O($^3$P) & \multicolumn{1}{c|}{22.3*} & 23.8 \cite{ref:Hikosaka}\\
O$_2^{2+}(X^1\Sigma_g^+)$ & O$^+$($^4$S)+O$^+$($^4$S) & \multicolumn{1}{c|}{31.6} & 32.4 \cite{ref:Magrakvelidze}\\ 
O$_2^{2+}(A^3\Sigma_u^+)$ & O$^+$($^4$S)+O$^+$($^4$S) & \multicolumn{1}{c|}{31.6} & 32.4 \cite{ref:Magrakvelidze}\\ 
O$_2^{2+}(1^5\Sigma_g^+)$ & O$^+$($^4$S)+O$^+$($^4$S) & \multicolumn{1}{c|}{31.6} & 32.4 \cite{ref:Magrakvelidze}\\ 
O$_2^{2+}(B^3\Pi_g)$ & O$^+$($^4$S)+O$^+$($^2$D) & \multicolumn{1}{c|}{35.0} & 35.7 \cite{ref:Magrakvelidze}\\  
O$_2^{2+}(C^3\Pi_u)$ & O$^+$($^4$S)+O$^+$($^2$D) & \multicolumn{1}{c|}{35.0} & 35.7 \cite{ref:Magrakvelidze}\\  
O$_2^{2+}(1^5\Pi_u)$ & O$^+$($^4$S)+O$^+$($^2$D) & \multicolumn{1}{c|}{35.0} & 35.7 \cite{ref:Magrakvelidze}\\  
O$_2^{2+}(W^3\Delta_u)$ & O$^+$($^4$S)+O$^+$($^2$D) & \multicolumn{1}{c|}{35.0} & 35.7 \cite{ref:Magrakvelidze}\\  
O$_2^{2+}(^5\Pi_g)$ & O$^+$($^4$S)+O$^+$($^2$D) & \multicolumn{1}{c|}{35.0} & 35.7 \cite{ref:Magrakvelidze}\\  
O$_2^{2+}(1^5\Sigma_u^-)$ & O$^+$($^4$S)+O$^+$($^2$D) & \multicolumn{1}{c|}{35.0} & 35.7 \cite{ref:Magrakvelidze}\\  
O$_2^{2+}(^3\Sigma_g^-)$ & O$^+$($^4$S)+O$^+$($^2$D) & \multicolumn{1}{c|}{35.0} & 35.7 \cite{ref:Magrakvelidze}\\  
O$_2^{2+}(2^3\Sigma_g^-)$ & O$^+$($^4$S)+O$^+$($^2$P) & \multicolumn{1}{c|}{36.6} & 37.3 \cite{ref:Magrakvelidze}\\
O$_2^{2+}(^3\Sigma_g^-)$ & O$^+$($^2$D)+O$^+$($^2$D)   & \multicolumn{1}{c|}{48.1*} & 39.0 \cite{ref:Magrakvelidze}\\
\end{tabular}
\caption{Atomic fragments resulting from the dissociation of O$_2$, O$_2^+$ and O$_2^{2+}$. We also provide the dissociation energy relevant to these atomic fragments, and compare with dissociation energies from Refs. \cite{ref:Magrakvelidze,ref:Hikosaka}. The  * denotes states calculated using only the CASSCF method due to lack of convergence of the MRCI method, as mentioned in Section \ref{ssec:PECs}.}
\label{table:DissFrags}
\end{table*}

\subsection{Sampling the initial conditions of the nuclei} \label{ssec:Wigner}
\begin{figure}
	\centering
    \includegraphics[width=0.48\textwidth]{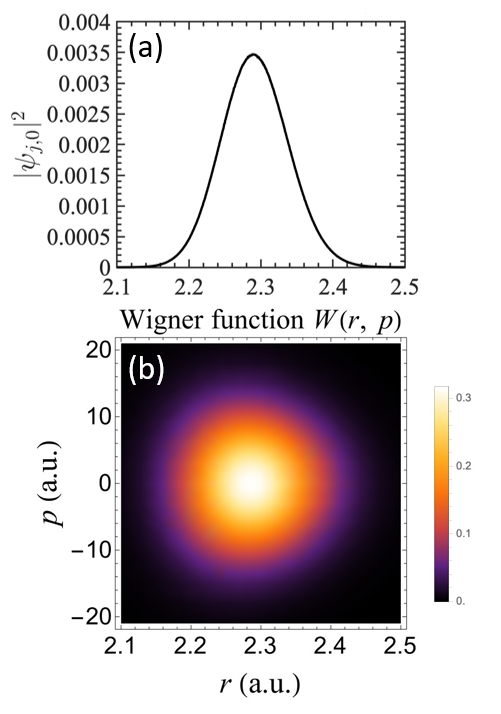}
\caption{Distributions used in importance sampling. Top: distribution of the internuclear distance using the square of the Morse wavefunction for the ground state of O$_2$. Bottom: distribution of the internuclear distance and the relative momenta using the Wigner function for the ground state of O$_2$. For each $r_0$ obtained using the distribution in \fig{fig:ImpSamp}(a), we sample $p_0$ using the distribution that is given by a cut of the Wigner function at $r_0$ in \fig{fig:ImpSamp}(b).}
\label{fig:ImpSamp}
\end{figure}

In the Monte-Carlo technique, at time step $n=0$, we need to specify  the initial internuclear distance, $r_0$, and relative momentum of the nuclei, $p_0=\mu v_0$. To avoid importance sampling of both $r_0$ and $p_0$ at the same time, which requires the use of a 2-dimensional distribution, we use an approximation and sample $r_0$ and $p_0$ as follows. First, for the internuclear distance, we employ importance sampling \cite{ref:ImpSamp} with the distribution given by the square of the Morse wavefunction \cite{ref:Frank} for the ground state of neutral O$_2$. The Morse potential is widely recognised as a good approximation for potentials of diatomic molecules  \cite{ref:Morse}. The  Morse potential is given by
\begin{equation}
U_M(r) = D_e[(1-e^{-\beta \abs{r - r_e}})^2-1],
\end{equation}
where $D_e$ is the dissociation energy of the ground state of O$_2$, which is equal to 5 eV. The coefficient $\beta$ is related to the frequency of vibration of the nuclei, $\omega_e$, by the following expression
\begin{equation}
\label{eqn:beta}
\omega_e = \beta\sqrt{\frac{2D_e}{\mu}},
\end{equation}
where $\mu$ is the reduced mass of the molecule. Using NIST, we assign $\omega_e$ to be equal to 1580.16 cm$^{-1}$ \cite{ref:Irikura}. Using \eq{eqn:beta} and the values we obtain for $D_e$, $\mu$ and $\omega_e$, we  compute $\beta$. The Morse wavefunction of the ground state of neutral O$_2$ \cite{ref:Frank} is given by
\begin{equation}
\begin{split}
\psi_{j,0}(r) &= N_{j,0}e^{-\xi/2}\xi^j L_0^{2j}(\xi)
\\
N_{j,0} &= \frac{\beta j}{\Gamma(2j+1)}
\\
\xi &= (2j+1)e^{-\beta \abs{r - r_e}},
\end{split}
\end{equation}
where $L_0^{2j}$ is an associated Laguerre polynomial \cite{ref:Slater} and $\Gamma$ is the Gamma function. The quantum number $j$ is related to the dissociation energy through
\begin{equation}
D_e = \frac{\beta^2}{2\mu}(j+\frac{1}{2})^2.
\end{equation}
Given $D_e$, $\mu$ and $\beta$, we solve for $j$. Next, after obtaining with importance sampling $r_0$, we use this $r_0$ as input to the Wigner function of the ground state of neutral O$_2$, which is given by
\begin{equation}
\label{eqn:Wigner}
W(\psi_{j,0}|r,p) = \frac{2}{\pi\Gamma(2j)}\xi^{2j}K_{-2ip/\beta}(\xi),
\end{equation}
where $K_{-2ip/\beta}(\xi)$ is the modified Bessel function of the third kind \cite{ref:Slater}. We sample the initial relative momentum $p_0$ by employing importance sampling with the Wigner function in \eq{eqn:Wigner} at $r=r_0$. In \fig{fig:ImpSamp}, we plot the distributions of the Morse wavefunction squared, $\abs{\psi_{j,0}(r)}^2$, and the Wigner function, $W(\psi_{j,0}|r,p)$, for the ground state of O$_2$. For any given $r_0$, the cut of the Wigner function at $r_0$ is a function of $p_0$ centred around 0 a.u., i.e. centred around the nuclei being at rest.

\subsection{Monte-Carlo technique} \label{ssec:MC}
In what follows, we outline the steps involved in the Monte-Carlo technique that describes the interaction between the O$_2$ molecule and the XUV pulse. 

For each event in the Monte-Carlo simulation, we start in the ground state of O$_2$ with an internuclear distance and relative momentum resulting from the sampling described in Section \ref{ssec:Wigner}. We choose the default time step to be $\Delta t=0.01$ fs, as we found convergence in our results at this granularity. For a given photon energy, at the start of a time step, we identify all the electronic transitions that are energetically accessible for a given photon energy from the molecular or atomic state $\alpha$ at this time to a state $i$ at the end of this time step. In order to determine which transition occurs in this time step, we calculate the transition rates for each single-photon ionization and Auger-Meitner process. The transition rates are given by
\begin{equation}
\label{eqn:AtomRates}
\begin{split}
\omega_{\alpha i}(t) &= \sigma_{\alpha i}J(t) \ \ \ \text{Photoionization}
\\
\omega_{\alpha i}(t) &= \Gamma_{\alpha i}\ \ \ \ \ \ \ \ \text{Auger-Meitner decay},
\end{split}
\end{equation}
where $\omega_{\alpha i}$ is the transition rate from the initial state $\alpha$ to the final state $i$, $\sigma_{\alpha i}$ is the photoionization cross section for this transition, $J(t)$ is the photon flux at time $t$ and $\Gamma_{\alpha i}$ is the Auger-Meitner rate. Note that we calculate these rates both for the molecular ions and for the atomic fragments that result from dissociation of the molecule due to the interaction with the XUV pulse. The molecular transition rates depend on the internuclear distance. The population of the molecular and atomic states follow an exponential decay law \cite{ref:Jurek}, i.e.
\begin{equation}
\label{eqn:DecayLaw}
P = P_0 e^{-\omega_{\alpha i}t},
\end{equation}
where $P, P_0$ are the populations of molecular and atomic states at times $t$ and at the start of a given time step, respectively. Hence, selecting a random value of $P$ such that $\frac{P}{P_0}\leq1$, the corresponding time for this transition for a given time step is given by
\begin{equation}
t_{\alpha i}(t) = -\frac{\log{\frac{P}{P_0}}}{\omega_{\alpha i}}.
\end{equation}
At the start of each time step, we compute the times $t_{\alpha i}$ for all transitions from the state $\alpha$ to energetically allowed states $i$. We identify the smallest time $t_{\alpha i}$, which corresponds to the most probable transition at a given time step. If this time is greater than the default time step of 0.01 fs, no electronic transition takes place and the time increases by the default time step of 0.01 fs. If $t_{\alpha i}$ is smaller than the default time step, then the time increases by $t_{\alpha i}$ and the transition to state $i$ takes place.

For a given time step, we propagate the nuclei, as described in Section \ref{ssec:VV}, using as the potential $U(r_n)$ and $U(r_{n+1})$, the potential of the state $\alpha$ at the distances corresponding to the start and the end of this time step, respectively. If a transition occurs, this time step is $t_{\alpha i}$, otherwise it is the default one, equal to 0.01 fs.

Finally, at each time step we check if dissociation occurs at the corresponding internuclear distance using the criteria outlined in Section \ref{ssec:Diss}. If dissociation does occur, from this time on in the Monte-Carlo simulation, we account for the interactions of the resulting atomic fragments with the XUV pulse.

\section{Experimental setup}\label{sec:Experiment}

The experiment of O$_2$ interacting with an XUV pulse was conducted at the reaction microscope (REMI) endstation \cite{ref:Meister,ref:Schmid} at the FEL in Hamburg, using the FLASH2 undulators \cite{ref:Ackermann,ref:Faatz}. The set-up allows one to analyze ionization and fragmentation processes and to measure the final ion states and their charge states in coincidence. In the ultra-high vacuum ($\approx$ 10$^{-11}$ mbar) detection chamber a supersonic gas jet is crossed at 90$^{\circ}$ with the focused XUV FEL beam. Electrons and ions which are generated during ionization are guided onto spatial and time sensitive detectors by means of collinear and homogenous electric and magnetic fields. Time-of-flight information and impact position of a particle enable the reconstruction of the momentum vectors of the particles at the instant of ionization. Momentum conservation between particles of the same molecule can be used to avoid false coincidences, e.g. between fragments of two molecules accidentally ionized in the same pulse.

For this experiment the FEL was operated in a pattern of 38 pulses temporally spaced by approximately 13 $\mu$s (77 kHz), repeating in a 10 Hz period. This results in an effective repetition rate of 380 Hz. Thanks to the variable gap undulators the photon energy was easily and repeatedly scanned between 20 eV and 42 eV in steps of 0.2 eV. Within this range the pulse energy, measured by the gas monitor detector (GMD) in the experimental hall, varied between 10 and 30 $\mu$J while about 30\% of this value is delivered on target. In Figure \ref{fig:KER_O+O+}, we plot the KER distribution of the atomic fragments in the O$^+$ + O$^+$ dissociation pathway obtained experimentally. In Figure \ref{fig:KER_O+O+}, we also compare the theoretical and experimental  results obtained in this work. As the FEL-pulse energy and other beam parameters change systematically with the photon energy, the experimental O$^+$ + O$^+$ ion yield in Figure \ref{fig:KER_O+O+}(b) is normalized with the simultaneously recorded H$_2^+$ yield from residual H$_2$ gas in the REMI chamber. This procedure is done by taking the total absorption cross section of H$_2$ into account \cite{ref:Samson}.

The electric field strength was set to about 16 Vcm$^{-1}$ such that Coulomb exploding O$^+$ pairs are detected in a $4\pi$ solid angle. The nozzle of the supersonic gas source was cooled to roughly -158 $^{\circ}$C, just above the condensation point of O$_2$ in order to prevent clogging. Cryogenic temperatures were chosen to reduce internal energy and to decrease momentum spread of the oxygen molecules in the jet.

\section{Results}\label{sec:Results}

In what follows, we present and discuss our results for the KER distributions of the atomic fragments in the O$^+$ + O$^+$ dissociation pathway of O$_2$ interacting with an XUV pulse. Here, the photon energy of the laser pulse ranges between 20 eV and 42 eV in increments of 1 eV for the theoretical results (0.2 eV for the experimental results). At each photon energy, we propagate in time $5\times10^6$ Monte-Carlo events. Concerning the XUV pulse, we consider laser intensities of $5\times 10^{12}$ W/cm$^2$ and $5\times 10^{14}$ W/cm$^2$ and full-width-at-half-maximum (FWHM) pulse durations of 50 fs and 100 fs. For each Monte-Carlo event, we propagate in time starting 500 fs before the peak of the laser pulse and ending 1000 fs afterwards. We have checked that our results converge when using these initial and final limits in the time propagation. At the end of the time propagation, we collect the events leading to the formation of the O$^+$ + O$^+$ pathway. For each of these events, we record the sequence of single-photon ionization processes. That is, we record the initial and final states involved in a photoionization transition, as well as the times and internuclear distances at which this transition occurs. We also record the final velocities of the two O$^+$ fragments  and calculate the sum of the kinetic energies of the atomic fragments to compare with experiment.

First, in Section \ref{ssec:PercentageMC} we discuss the probability out of all  Monte-Carlo events to obtain the O$^+$ + O$^+$ pathway as a function of photon energy. Then, in Section \ref{ssec:KER}, we plot the KER distribution of the two O$^+$ fragments for all O$^+$ + O$^+$ events as a function of photon energy. We compare our theoretical results to the experimental KER distribution and identify the main ionization sequences that lead to the formation of the O$^+$ + O$^+$ pathway. In Section \ref{ssec:Intensity}, we plot the KER distribution for a higher intensity of the laser pulse in order to understand the effect of intensity on the kinetic-energy spectra of the O$^+$ fragments. Finally, in Section \ref{ssec:Dists}, we plot the distribution of internuclear distances at which single-photon ionization occurs. We use this distribution, as well as the various sequences of ionization processes, in order to explain the main features of the KER distribution.

Note that in our theoretical calculations, we do not include excited states of O$_2^+$ and O$_2^{2+}$. For the photon energies we consider in this work, it is energetically allowed to transition from the ground state of O$_2$ to excited states of O$_2^+$ and also to transition from O$_2^+$ states to excited states of O$_2^{2+}$. However, such transitions involve the calculation of matrix elements where in the final state there is a simultaneous one-electron excitation and one-electron ionization. For instance, to transition to the excited state O$_2^+(f^4\Pi_g)$ from the ground state of O$_2$ would involve the ionization of an electron from a $\pi_u$ orbital and the excitation of another electron from a $\pi_u$ to a $\pi_g$ orbital. Hence, the accurate computation of such matrix elements is only possible when electron-electron correlation is included in the description of the wavefunctions. This is not accounted for in our current formulation, where we utilise Hartree-Fock wavefunctions. Moreover, the reason that it is a reasonable approximation to not include these excited states is that from the O$_2^+$  excited states, one could transition to non-excited states of O$_2^{2+}$ by single-photon absorption. However, in our calculations, we do get these same non-excited states of O$_2^{2+}$ by transitions from non-excited states of O$_2^+$. Finally, the O$_2^{2+}$ excited states, accessible by the photon energies in this calculation, tunnel to non-excited states of O$_2^{2+}$ due to coupling effects between states. These final non-excited states of O$_2^{2+}$ we do access in our calculation, resulting in similar KER spectra as in the experiment.

\subsection{Probability of the O$^+$ + O$^+$ dissociation pathway for different pulse durations}\label{ssec:PercentageMC}

\begin{figure}[H]
	\centering
	\includegraphics[width=0.48\textwidth]{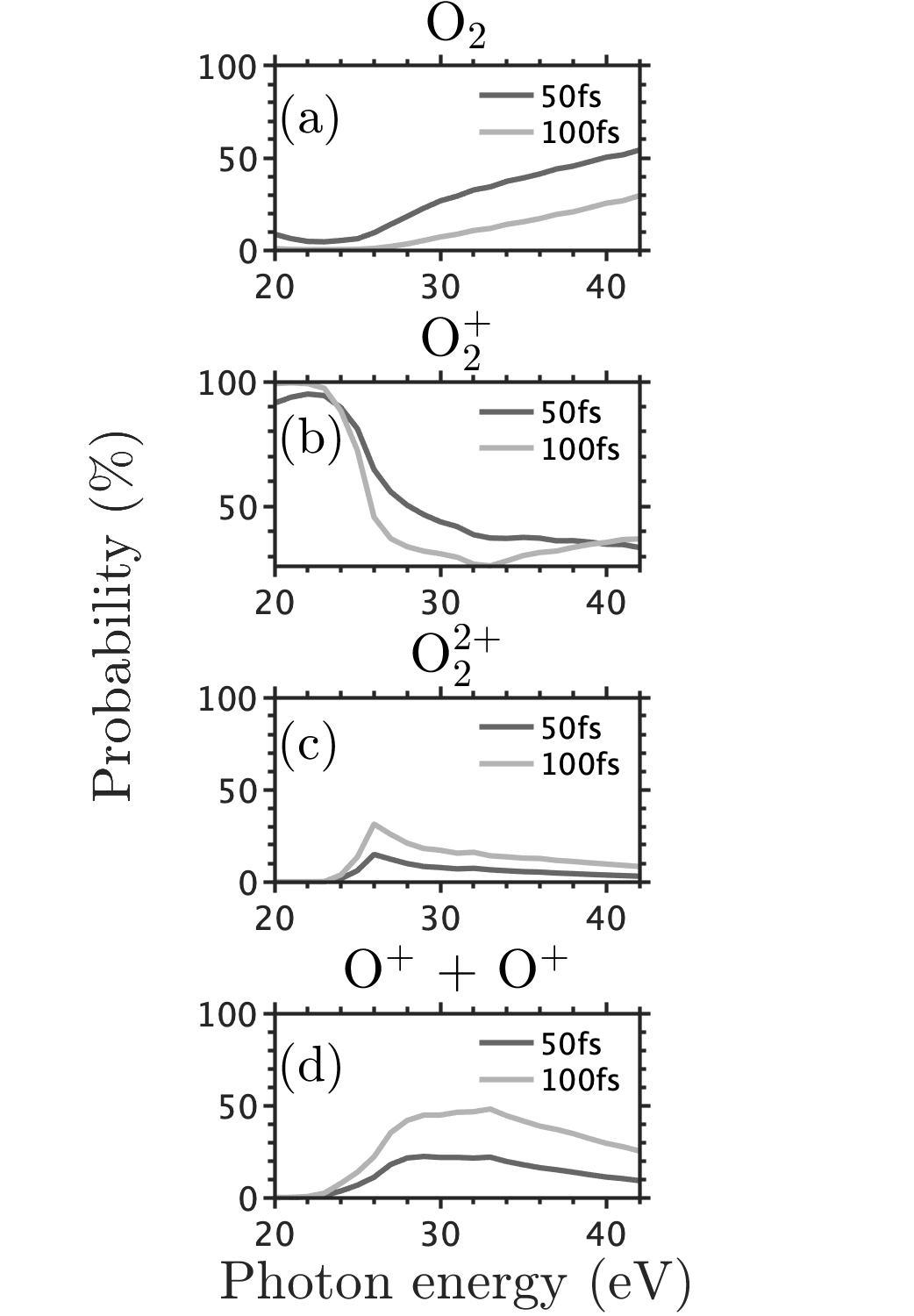}
\caption{Probability, out of all Monte-Carlo events, of the (a)  O$_2$, (b) O$_2^+$, (c) O$_2^{2+}$ pathways and of the (d) O$^+$ + O$^+$ dissociation pathway as a function of photon energy. The intensity of the laser pulse is $5\times 10^{12}$ W/cm$^2$ and the FWHM are 50 fs and 100 fs.}
\label{fig:Percentages}	
\end{figure}

\begin{figure}[H]
	\centering
	\includegraphics[width=0.49\textwidth]{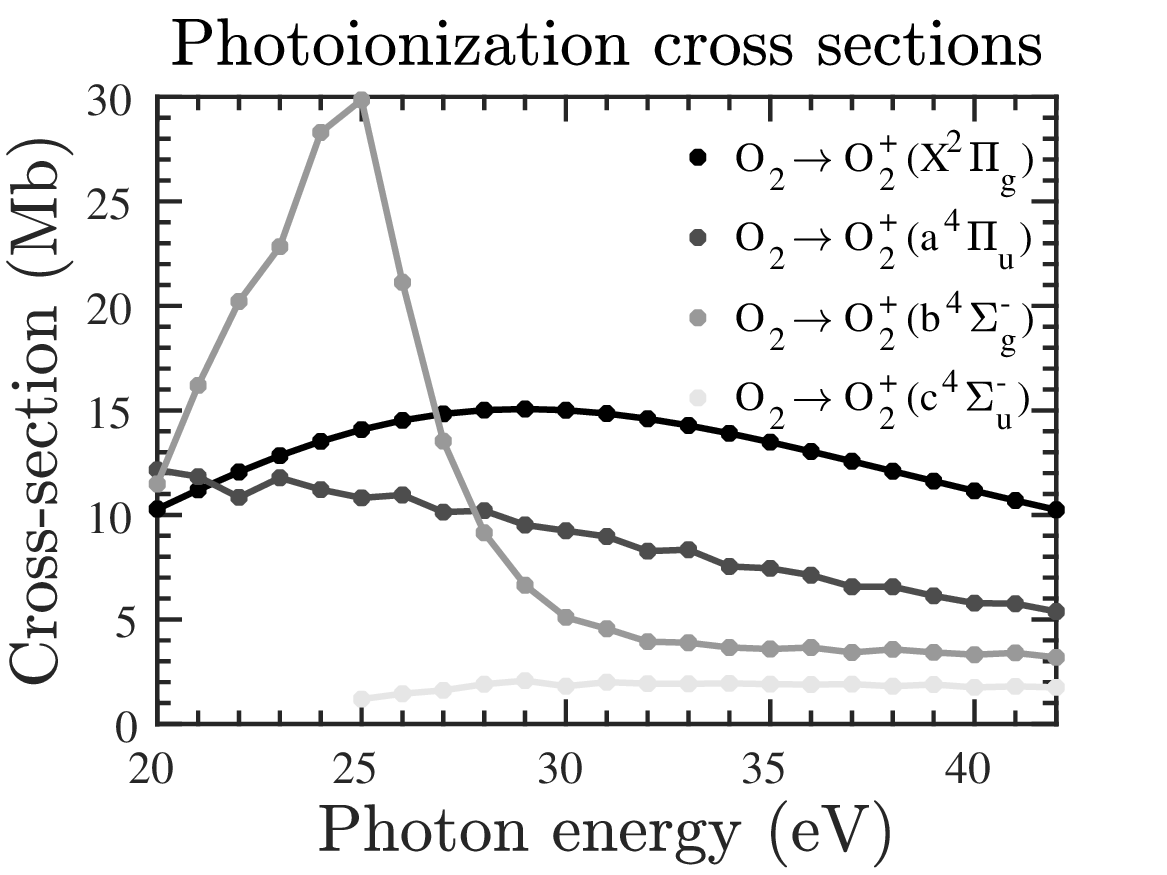}
\caption{Photoionization cross sections to transition from O$_2$ to O$_2^+$ at the equilibrium distance of O$_2$, $r_e=2.28$ a.u., as a function of photon energy.}
\label{fig:Xsections}	
\end{figure}

In \fig{fig:Percentages}, we plot, out of all Monte-Carlo events, the probability of the O$_2$, O$_2^+$ and O$_2^{2+}$, i.e. of the non-dissociating pathways, as well as of  the O$^+$ + O$^+$ dissociation pathway as a function of photon energy. We use a laser-pulse intensity of $5\times 10^{12}$ W/cm$^2$ and FWHM durations of 50 fs and 100 fs. We find that it is more probable to obtain the O$^+$ + O$^+$ pathway for a FWHM of 100 fs versus a 50 fs one. This is expected, since for a longer laser pulse there is more of a chance to absorb photons. This is also consistent with the probability of O$_2^{2+}$ being higher, while the probability for O$_2$ is lower for 100 fs. We also find that for both pulse durations, the probability of the O$^+$ + O$^+$ pathway is very small for small photon energies, it peaks at roughly 30 eV photon energy and then decreases with increasing photon energy. For small photon energies, the  probability of the O$^+$ + O$^+$ pathway is small, even though the probability of O$_2^+$ formation is large, since no transition is allowed from O$_2^+$ to O$_2^{2+}$. For larger photon energies, the probability of the O$^+$ + O$^+$ pathway decreases since the single-photon ionization cross sections for transitioning from O$_2$ to O$_2^+$ states decrease or remain constant for these photon energies. This is shown in \fig{fig:Xsections}, where for photon energies above 30 eV for transitions to the O$_2^+$ states X$^2\Pi_g$ and a$^4\Pi_u$ the cross sections decrease, while they remain flat for transitions to the states b$^4\Sigma_g^-$ and c$^4\Sigma_u^-$.

\subsection{KER distribution and the main ionization sequences} \label{ssec:KER}

\begin{figure}
	\centering
    \includegraphics[width=0.48\textwidth]{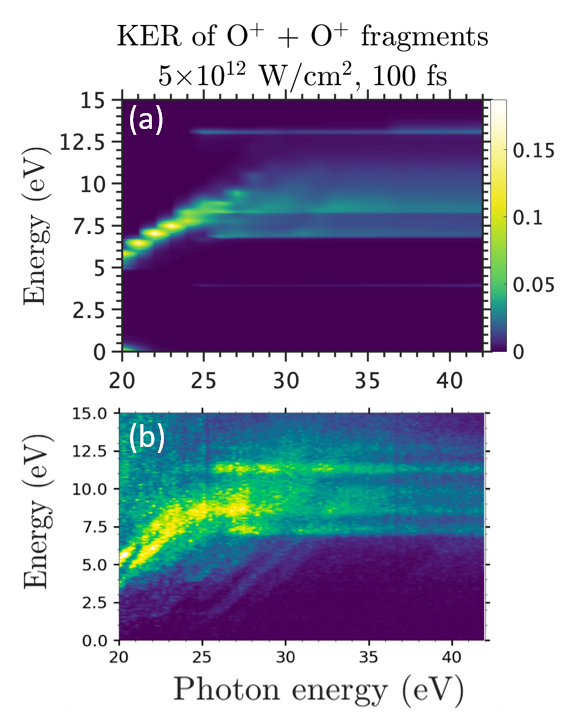}
\caption{Kinetic-energy-release spectra of the O$^+$ + O$^+$ pathway as a function of photon energy. (a) Theoretical results obtained in the current work using a laser-pulse intensity equal to $5\times 10^{12}$ W/cm$^2$ and FWHM of 100 fs. (b) Experimental results, for more information on the set-up see Ref.\cite{ref:Corlin,ref:Schmid}.}
\label{fig:KER_O+O+}
\end{figure}

In \fig{fig:KER_O+O+}(a), we plot the sum of the kinetic energies of the atomic fragments of the O$^+$ + O$^+$ dissociation pathway as a function of the photon energy. At each photon energy, the KER distribution is normalized to 1, i.e. we divide by all  O$^+$ + O$^+$ events at this photon energy. For our simulations, we use a weak laser-pulse intensity of $5\times 10^{12}$ W/cm$^2$ and a FWHM of 100 fs in order to closely resemble the parameters of the laser pulses used in the experiment. We compare our theoretical results for the KER distribution, shown in \fig{fig:KER_O+O+}(a), with the experimental ones shown in \fig{fig:KER_O+O+}(b). We find that our theoretical results reproduce well most of the features in the experimental KER distribution. That is, the KER distribution peaks at 5 eV for 20 eV photon energy, increases with increasing photon energy, reaching 8 eV at 25 eV photon energy. From 25 eV photon energy onwards, we observe numerous peaks in the kinetic-energy spectra, ranging from 7 eV to 13 eV. These peaks in the spectra remain mostly constant from 25 eV until 42 eV photon energy, giving rise to the straight lines that we see in \fig{fig:KER_O+O+}(a) and \fig{fig:KER_O+O+}(b). However, our theoretical results do not reproduce the experimental KER peak at approximately 11 eV. It is possible that this is due to the exclusion of the excited states in our calculations.

\begin{figure}
	\centering
    \includegraphics[width=0.49\textwidth]{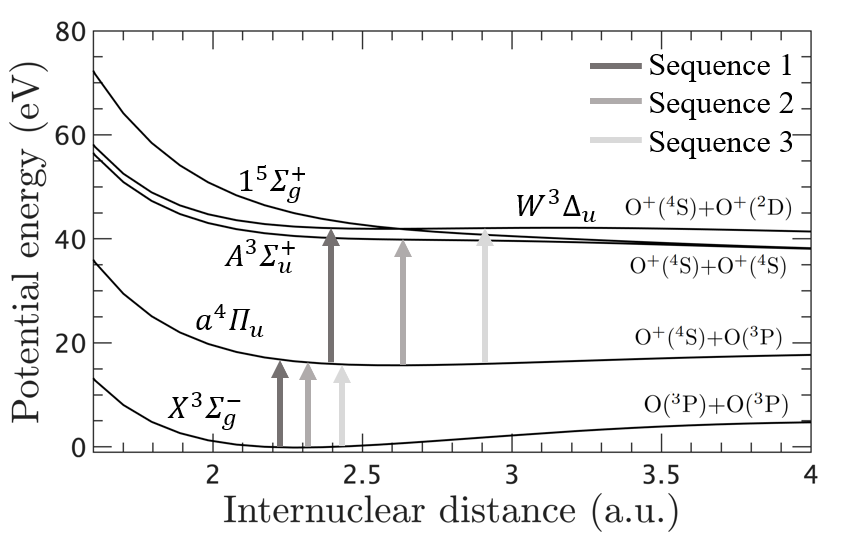}
\caption{Schematic depiction of the two single-photon ionization transitions, first, from the ground state to an O$_{2}^{+}$ state and, then, from the O$_{2}^{+}$ state  to an O$_{2}^{2+}$ state for ionization sequences 1,2,3. }
\label{fig:Annotated_PEC}
\end{figure}

To explain the features of the KER distribution of the O$^+$ + O$^+$ pathway, in Table \ref{table:Pathways}, we identify the main ionization sequences leading to the formation of two O$^+$ fragments. Each ionization sequence involves a single-photon ionization leading to a transition from the ground state of O$_2$, i.e. the O$_2(X^3\Sigma_g^-)$ state, to a O$_2^+$ state and a subsequent single-photon ionization leading to a transition from an O$_2^+$ state to an O$_2^{2+}$ state. The difference between these sequences are the O$_2^+$ and O$_2^{2+}$ states involved in the photoionization transitions. After transitioning to an O$_2^{2+}$ state, as described in Section \ref{ssec:Diss}, this  state dissociates to two O$^{+}$ fragments, shown in Table \ref{table:DissFrags}. As shown in \fig{fig:PECs}, the PECs of all the O$_2^{2+}$ states involved in the ionization sequences 1 through 8 in Table \ref{table:Pathways} are repulsive due to the Coulomb potential of the atomic ions. As a result, during time propagation the internuclear distance increases rapidly leading to the formation of two O$^+$ fragments. For additional clarity, in \fig{fig:Annotated_PEC}, we schematically depict the two single-photon ionization transitions involved for the ionization sequences 1,2,3.

 As mentioned earlier, we do not include excited states of O$_2^{2+}$,  such as the $1^1\Pi_g$, $1^1\Delta_u$, $1^1\Sigma_u^-$ and $B^3\Sigma_u^-$ states. However, these states are coupled to non-excited states of O$_2^{2+}$, namely the $C^3\Pi_u$, $1^5\Sigma_g^+$ and $1^5\Pi_u$ states, all three of which are the final O$_2^{2+}$ states for some of the eight ionization sequences in Table \ref{table:Pathways}. This is the reason the theoretical KER distribution in \fig{fig:KER_O+O+}(a) still reproduces the features that appear in the experimental distribution in \fig{fig:KER_O+O+}(b).
\begin{table}[t]
\begin{tabular}{c|l}
 & Ionization Sequence \\ \hline
1 & O$_2 \rightarrow \text{O}_2^+(a^4\Pi_u) \rightarrow \text{O}_2^{2+}(1^5\Sigma_g^+)$ \\
2 & O$_2 \rightarrow \text{O}_2^+(a^4\Pi_u) \rightarrow \text{O}_2^{2+}(A^3\Sigma_u^+)$\\
3 & O$_2 \rightarrow \text{O}_2^+(c^4\Sigma_u^-) \rightarrow \text{O}_2^{2+}(C^3\Pi_u)$\\
4 & O$_2 \rightarrow \text{O}_2^+(a^4\Pi_u) \rightarrow \text{O}_2^{2+}(W^3\Delta_u)$\\
5 & O$_2 \rightarrow \text{O}_2^+(b^4\Sigma_g^-) \rightarrow \text{O}_2^{2+}(1^5\Pi_u)$\\
6 & O$_2 \rightarrow \text{O}_2^+(a^4\Pi_u) \rightarrow \text{O}_2^{2+}(1^5\Pi_u)$\\
7 & O$_2 \rightarrow \text{O}_2^+(X^2\Pi_g) \rightarrow \text{O}_2^{2+}(W^3\Delta_u)$\\
8 & O$_2 \rightarrow \text{O}_2^+(X^2\Pi_g) \rightarrow \text{O}_2^{2+}(A^3\Sigma_u^+)$
\end{tabular}
\caption{Main ionization sequences leading to the formation of the O$^+$ + O$^+$ dissociation pathway. These eight ionization sequences account for almost all O$^+$ + O$^+$ events.}
\label{table:Pathways}
\end{table}

\subsection{KER dependence on intensity of the laser pulse} \label{ssec:Intensity}

\begin{figure}
	\centering
    \includegraphics[width=0.46\textwidth]{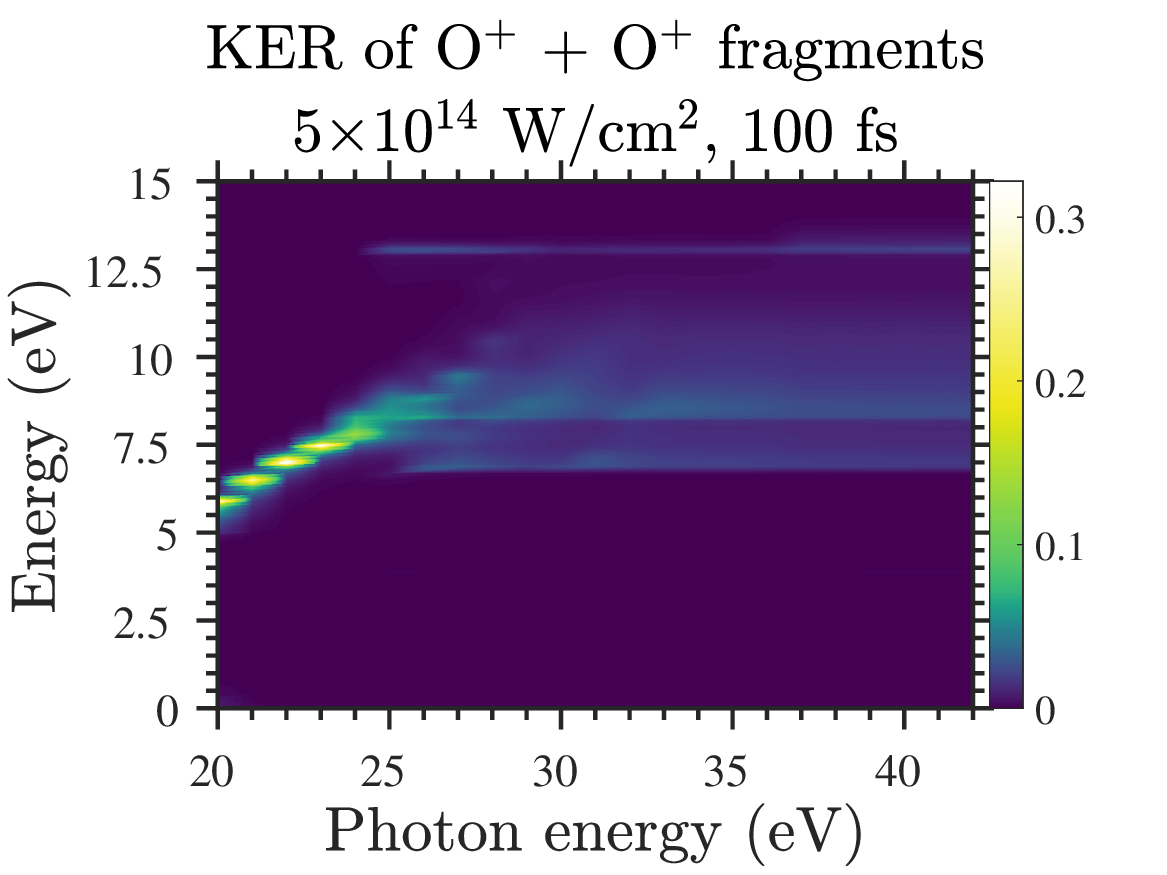}
\caption{Same as \fig{fig:KER_O+O+}(a) with a laser intensity of $5\times 10^{14}$ W/cm$^2$.}
\label{fig:KER_O+O+_2}
\end{figure}

\begin{figure}
	\centering
	\includegraphics[width=0.48\textwidth]{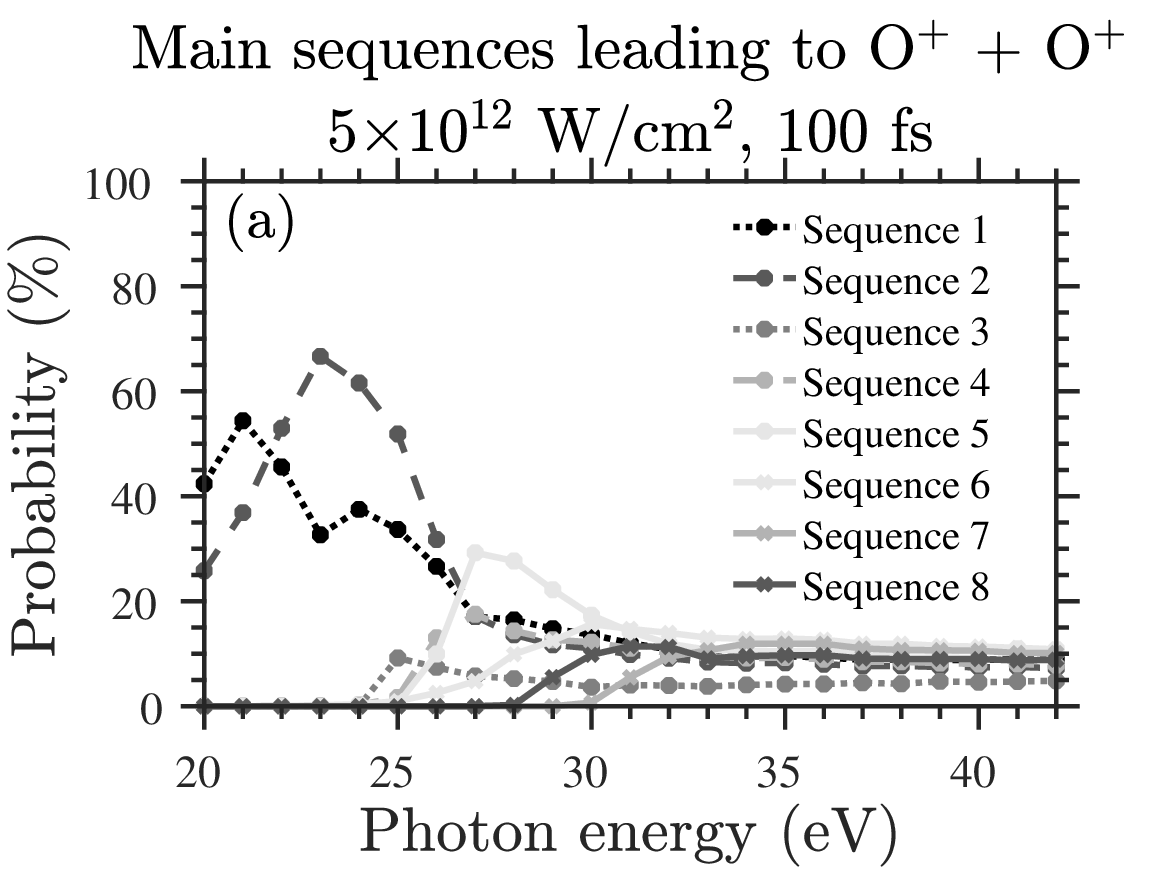}
	\includegraphics[width=0.48\textwidth]{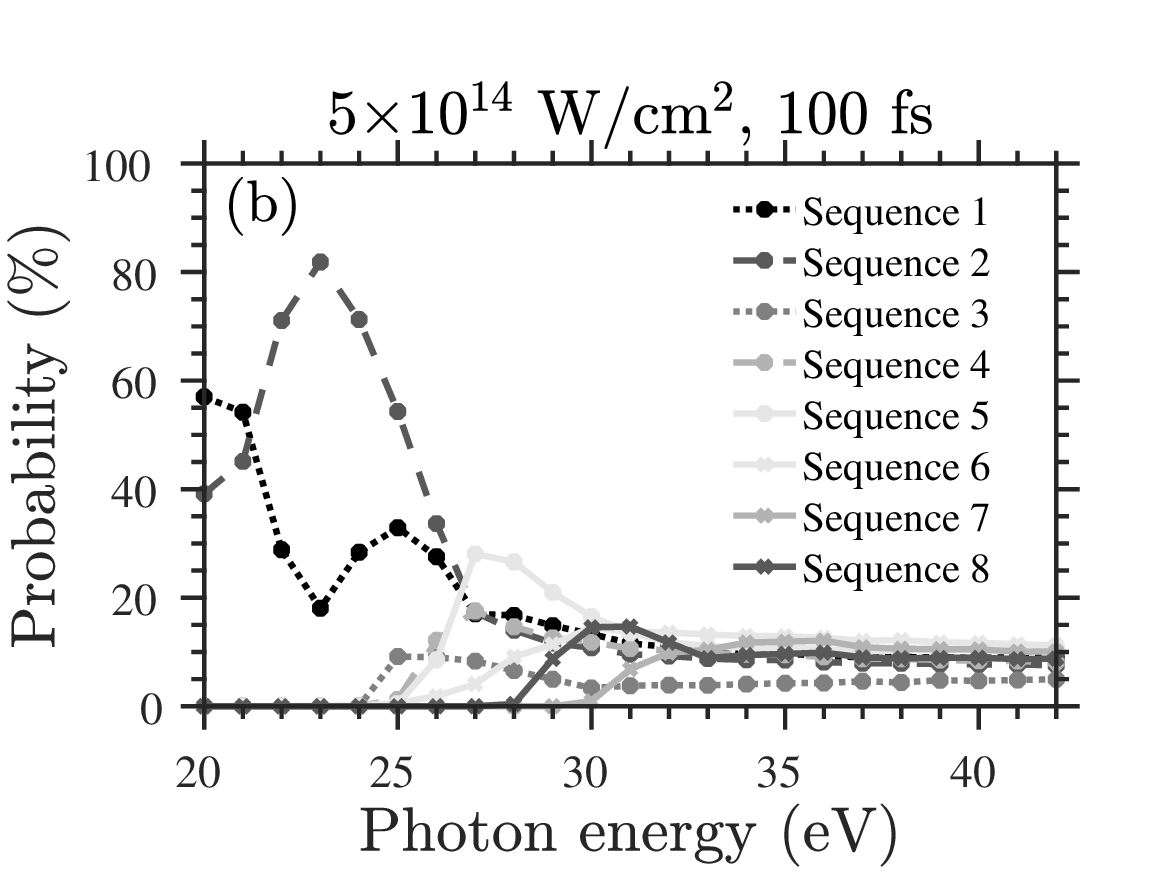}
\caption{Probability, computed out of all O$^+$ + O$^+$ events, of each of the 8 main ionization sequences leading to the formation of the O$^+$ + O$^+$ pathway, as a function of the photon energy. The FWHM of the laser pulse is 100 fs and the laser intensity is (a) $5\times 10^{12}$ W/cm$^2$ (b) $5\times 10^{14}$ W/cm$^2$.}
\label{fig:Pathways}	
\end{figure}

In what follows, we investigate how the KER distribution of the O$^+$ + O$^+$ pathway changes with the laser-pulse intensity. In \fig{fig:KER_O+O+_2}, we plot the KER distribution for a laser-pulse intensity of $5\times 10^{14}$ W/cm$^2$, while in \fig{fig:KER_O+O+}(a) the laser-pulse intensity is smaller and equal to $5\times 10^{12}$ W/cm$^2$. Comparing the KER distribution in \fig{fig:KER_O+O+_2} with \fig{fig:KER_O+O+}(a), we find that the KER distribution for each photon energy from 20 eV to 25 eV is broader for the lower intensity. From 25 eV photon energy onwards, the KER distributions are similar in \fig{fig:KER_O+O+_2} and \fig{fig:KER_O+O+}(a). To explain 
 the wider KER distributions for small photon energies for the lower intensity, in \fig{fig:Pathways}, we plot the probabilities of each of the eight ionization sequences as a function of photon energy. We find that for photon energies roughly up to 25 eV, only sequences 1 and  2 contribute to the O$^+$ + O$^+$ pathway. For the higher intensity, sequence 2 contributes more, while for the lower intensity both sequences contribute roughly equally. The contribution of one versus two ionization sequences is consistent with a narrower KER distribution for the higher intensity.   For 25 eV photon energy onwards, we find that almost all ionization sequences contribute roughly the same for both intensities, resulting in similar KER distributions for these photon energies for both intensities.

\subsection{Main features of the KER distribution}\label{ssec:Dists}

\begin{figure*}
	\centering
    \includegraphics[width=1\textwidth]{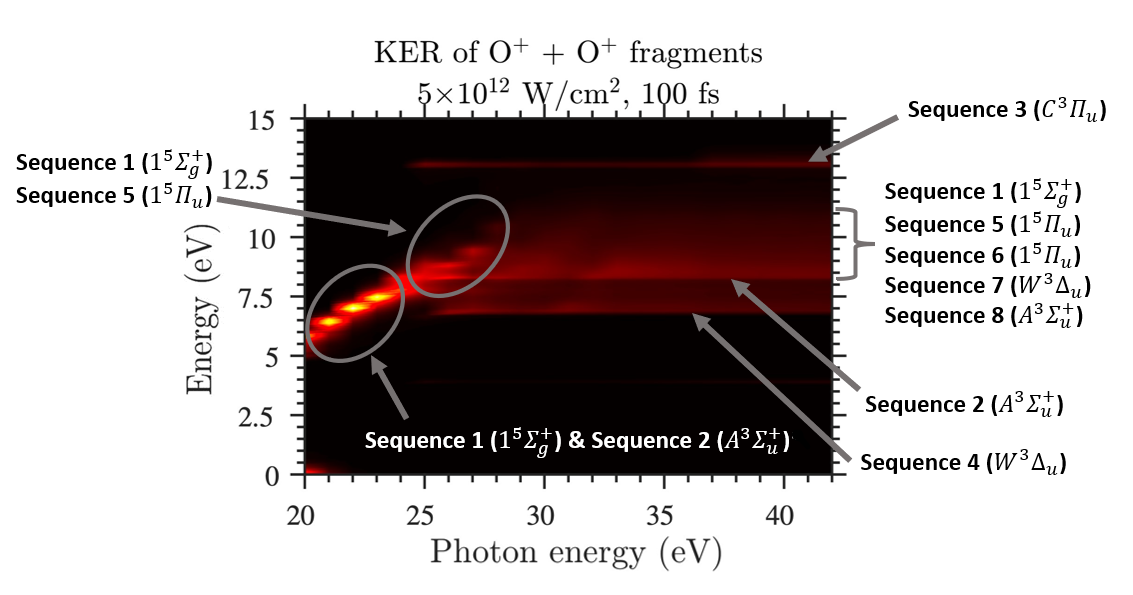}
\caption{Annotated kinetic-energy release spectra of the atomic fragments in the O$^+$ + O$^+$ dissociation pathway, associating  different ionization sequences to different features. The laser-pulse intensity is equal to $5\times 10^{12}$ W/cm$^2$ and the FWHM is equal to 100 fs.}
\label{fig:Annotated_KER}
\end{figure*}

In this section, we associate the main features of the KER distribution as a function of photon energy with the ionization sequences. Plotting the KER distribution of each sequence individually (not shown), we are able to assign features from the overall KER distribution to specific sequences. This correspondence is shown in \fig{fig:Annotated_KER}, for the KER distribution of the O$^+$ + O$^+$ pathway for the laser-pulse intensity of $5\times 10^{12}$ W/cm$^2$ and FWHM of 100 fs. For photon energies from 20 eV to 25 eV, the centre of the KER distributions changes almost linearly from 5 eV to 9 eV. This is due to the ionization sequences 1 and 2, as we have previously mentioned in Section \ref{ssec:Intensity}. Another aspect of the KER distributions is an almost linear increase from 9 eV to 11 eV between 25 eV and 28 eV photon energy. We find that this KER feature is due to the ionization sequences 1 and 5. Also, for photon energies higher than 25 eV, we see a series of constant lines in the kinetic-energy spectra. The  line corresponding to the highest kinetic energy of  13 eV  is due to sequence 3, while the line corresponding to the lowest kinetic energy of 7 eV  is due to sequence 4. The sequences 1, 5, 6, 7 and 8 equally contribute to the broad spectra between 8 eV and 11 eV kinetic energy.

\begin{figure*}
	\centering
    \includegraphics[width=1\textwidth]{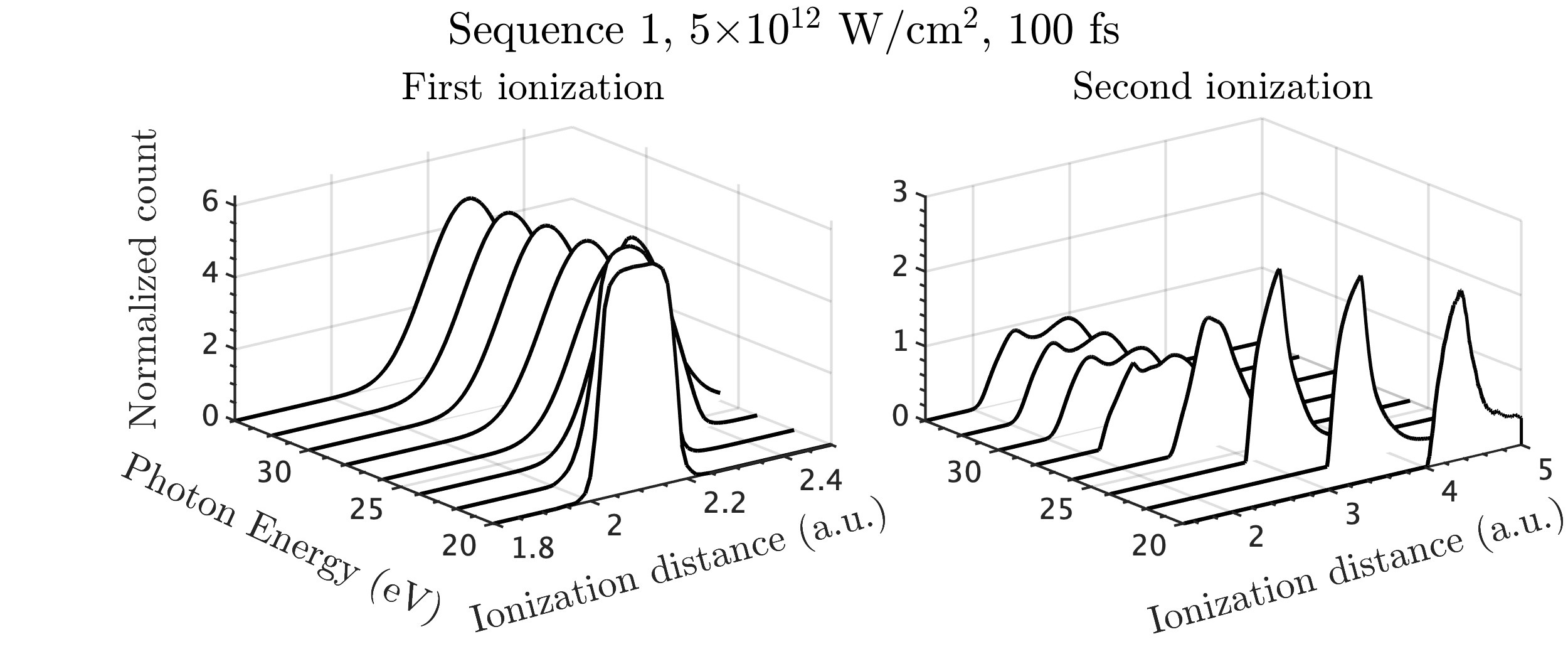}
\caption{For ionization sequence 1, distributions of internuclear distances when the first single-photon ionization (left column) and second single-photon ionization (right column) take place.}
\label{fig:Distributions}
\end{figure*}

To explain the  features of \fig{fig:Annotated_KER}, in \fig{fig:Distributions}, we plot the distributions of the internuclear distances when each of the two single-photon ionization transitions occur for ionization sequence 1. The reason we focus on this sequence is that we obtain similar results for all eight sequences. In \fig{fig:Distributions}, for the first ionization process, we see that the distribution of internuclear distances is narrower for small compared to larger photon energies. The reason is that as we increase the photon energy the first ionization step from the ground O$_2$ state to the single ionized O$_2^+$ state is energetically accessible at more internuclear distances. In addition, we find that for all photon energies, the distribution of internuclear distances are centred around the equilibrium distance of O$_2$, which is $r_e=2.28$ a.u. For the second ionization transition, we find that the distribution of internuclear distances is narrower for small photon energies compared to photon energies above 25 eV. Also, the second ionization transition takes place at large internuclear distances, around 4.5 a.u., for 20 eV photon energy, decreasing to roughly 3 a.u. at 26 eV photon energy. For photon energies above 26 eV, the distribution of internuclear distances doesn't change and is centred around small internuclear distances of roughly 2.5 a.u. The reason is that for small photon energies, a transition from an O$_2^+$ state to an O$_2^{2+}$ state is only allowed for larger distances, note from \fig{fig:PECs} the energy difference between the PECs of O$_2^+(a^4\Pi_u)$ and O$_2^{2+}(1^5\Sigma_g^+)$ involved in sequence 1. 

For photon energies between 20 eV and 25 eV, the decrease of the internuclear distance in the second ionization step is consistent with the linear increase in the centre of the KER distribution in \fig{fig:Annotated_KER}. Indeed, in \fig{fig:PECs}, we see that at larger internuclear distances the PEC of the O$_2^{2+}(1^5\Sigma_g^+)$ state involved in sequence 1 is shallower and hence the derivative of the PEC, which is the repulsive force, is smaller leading to a smaller velocity gain of the two O$^+$ fragments at dissociation. Also, the spectral lines remaining constant for photon energies higher than 25 eV is consistent with the width of the distribution of internuclear distances being the same for higher photon energies in \fig{fig:Distributions}. The value of the sum of the final kinetic energies of the O$^+$ atomic fragments, and hence the kinetic energy  each spectral line corresponds to, depends on the slope of the PECs of the O$_2^{2+}$ states at the internuclear distance when the second ionization transition takes place.

\section{Conclusion}
We have presented a hybrid quantum-classical technique to account for both the electronic structure and electron escape as well as the nuclear dynamics of a diatomic molecule during its interaction with an XUV laser pulse. In our technique, we treat quantum mechanically the electronic structure and ionization as well as the single-photon ionization and Auger-Meitner processes. In addition, we compute, with accurate quantum-chemistry methods, the potential-energy curves for molecular ion states of O$_2$ up to O$_2^{2+}$. We then use these potential-energy curves to compute the force between the two nuclei and classically account for the final velocities of the atomic fragments. Both the quantum and classical aspects of our techniques are incorporated in a stochastic Monte-Carlo calculation that accounts for the interaction of an O$_2$ molecule with an XUV pulse. The accuracy of our technique is demonstrated by comparing the sum of the kinetic energies of the two O$^+$ atomic fragments in the O$^+$ + O$^+$ pathway as a function of photon energy with experimental results. We find very good agreement with experiment. Moreover, we are able to associate the main features of the kinetic-energy release distribution as a function of photon energy for the O$^+$ + O$^+$ pathway to the main ionization sequences leading to this pathway. Our technique is general and can be applied to any diatomic molecule.

\section{Acknowledgements}
The authors A. E. and M. M. acknowledge the use of the UCL Myriad High Throughput Computing Facility (Myriad@UCL), and associated support services, in the completion of this work. A. E. acknowledges the Leverhulme Trust Research Project Grant No. 2017-376. M. M. acknowledges funding from the EPSRC project 2419551. The team at beamline FL26 of FLASH2 and the operators of FLASH at DESY are gratefully acknowledged. 

\bibliography{O2KERPaper}{}

\end{document}